# On Some Manipulations with Fuzzy Processes

**Lucian Luca, Lucian L. Luca,**
**„Tibiscus" University, Timişoara, Romania**

**ABSTRACT.** The paper starts from the observation on the complexity of the manipulation of fuzzy processes that increases very rapidly with the extents of the processes representation. Therefore, a productive approach is to divide the problem into smaller parts, treated separately and then the results combined. Some algebraic results obtained by the authors are presented**.**
**KEYWORDS:** fuzzy process, refination, robustness, robust process, chaotic process

## Introduction

We remind the notion of *fuzzy process* that we introduced in [LD01], a formalism for the notion of fuzzy contract between a device and its environment. Such a contract specifies the device-environment interface in terms of executions, which can be sequences of events, time functions, etc; yet we will consider them justly as elements of an arbitrary set $E$.

Let $E$ be the set of all executions and $\Delta : E \to [0,1]$ and $\Gamma : E \to [0,1]$ be two fuzzy subsets of E. In what follows, we note with:
$X = \{x \in E \mid \Delta(x) > 0\}$, $Y = \{x \in E \mid \Gamma(x) > 0\}$, $B = \{x \in E \mid \Delta(x) = \Gamma(x) = 0\}$
and we respectively call:
- $X$ – the set of accessible executions;
- $Y$ – the set of acceptable executions;
- $B$ – the set of rejections.

Additionally, we note $\Delta_X = \Delta_{/X}$, $\Gamma_Y = \Gamma_{/Y}$

**Definition 1:** *The pair $p = (\Delta_X, \Gamma_Y)$, where $\Delta_X$ and $\Gamma_Y$ are defined as above, is called a (vague) fuzzy process over E.*

*The set of all fuzzy processes over a pair of crisp subsets X and Y of E, as above, is called the space of the fuzzy process of (X, Y), and the set of all fuzzy process over E is called the space of the fuzzy process of E.* ∎

In [Luc03] we studied in detail the refination ($\sqsubseteq$), we defined and studied the operations with fuzzy processes: the sum ($\oplus$), the product ($\otimes$), the intersection ($\sqcap$), the reunion ($\sqcup$), and the reflection (-).

As we could notice, the complexity of manipulation of fuzzy processes increases very rapidly with the representation of their extent. Therefore, a productive approach consists in dividing the problem into smaller parts, treated separately and then combining the results.

## 1 A new definition of refination

We start by presenting some algebraic results we have obtained:

**Proposition 1**: *Let us have three fuzzy processes p, q and r over the E set of executions, then*

$$p \sqsubseteq q \;\Rightarrow\; p \otimes r \sqsubseteq q \otimes r$$

***Demonstration.***

$$p \sqsubseteq q \;\Leftrightarrow\; \begin{cases} \Delta^p(x) \geq \Delta^q(x) \\ \Gamma^p(x) \leq \Gamma^q(x) \end{cases} \Rightarrow \begin{cases} X_p \supseteq X_q \\ Y_p \subseteq Y_q \end{cases}$$

$$\Downarrow$$

$$p \otimes r \sqsubseteq q \otimes r \;\Leftrightarrow\; \begin{cases} \Delta^{p \otimes r}(x) \geq \Delta^{q \otimes r}(x) \\ \Gamma^{p \otimes r}(x) \leq \Gamma^{q \otimes r}(x) \end{cases} \Rightarrow \begin{cases} X_{p \otimes r} \supseteq X_{q \otimes r} \\ Y_{p \otimes r} \subseteq Y_{q \otimes r} \end{cases}$$

Yet,

$$\Delta^{p \otimes r}_{X_{p \otimes r}}(x) = \min_{x \in X_p \cap X_r} \{\Delta^p(x), \Delta^r(x)\}$$

$$\Delta^{q \otimes r}_{X_{q \otimes r}}(x) = \min_{x \in X_q \cap X_r} \{\Delta^q(x), \Delta^r(x)\}$$

$$\begin{cases} X_{p \otimes r} = X_p \cap X_r \\ X_{q \otimes r} = X_q \cap X_r \\ X_p \supseteq X_q \end{cases} \Rightarrow X_{p \otimes r} \supseteq X_{q \otimes r}$$

$$Y_{p \otimes r} = (Y_p \cap Y_r) \cup (\tilde{X}_p \cap \tilde{Y}_r) \cup (\tilde{Y}_p \cap \tilde{X}_r)$$
$$Y_{q \otimes r} = (Y_q \cap Y_r) \cup (\tilde{X}_q \cap \tilde{Y}_r) \cup (\tilde{Y}_q \cap \tilde{X}_r)$$
$$Y_p \subseteq Y_q \quad \Rightarrow \quad Y_p \cap Y_r \subseteq Y_p \cap Y_r$$
$$\tilde{X}_p \subseteq \tilde{X}_q \quad \Rightarrow \quad \tilde{X}_p \cap \tilde{Y}_r \subseteq \tilde{X}_q \cap \tilde{Y}_r$$
$$\tilde{Y}_q \subseteq \tilde{Y}_p \quad \Rightarrow \quad \tilde{Y}_q \cap \tilde{X}_r \subseteq \tilde{Y}_p \cap \tilde{X}_r$$

that is $Y_{p \otimes r} \subseteq Y_{q \otimes r}$ ∎

**Corollary 1:** *Let the fuzzy processes p and q be over the E set of executions, then* $\quad p \sqsubseteq q \quad \Rightarrow \quad p \sqsubseteq p \otimes q$

***Demonstration*:** *Considering from the proposition 1 we obtain:*
$$p \sqsubseteq q \quad \Rightarrow \quad p = p \otimes p \sqsubseteq q \otimes p = p \otimes q \quad \blacksquare$$

**Corollary 2:** *Let the fuzzy processes $p_1$, $p_2$, $q_1$, $q_2$ and q be over the E set of executions:*

   *i)* $\quad p_1 \sqsubseteq q_1 \land p_2 \sqsubseteq q_2 \Rightarrow p_1 \otimes p_2 \sqsubseteq q_1 \otimes q_2$
   *ii)* $\quad p_1 \sqsubseteq q \land p_2 \sqsubseteq q \Rightarrow p_1 \otimes p_2 \sqsubseteq q$

***Demonstration.*** *i) From proposition 1 we obtain that:*
$$p_1 \sqsubseteq q_1 \Rightarrow p_1 \otimes p_2 \sqsubseteq q_1 \otimes p_2$$
$$p_2 \sqsubseteq q_2 \Rightarrow p_2 \otimes q_1 \sqsubseteq q_2 \otimes q_1 \Rightarrow$$
$$q_1 \otimes p_2 \sqsubseteq q_1 \otimes q_2 \quad \text{(from the commutativity of } \otimes\text{)}$$

*From the transitivity of the refination relation:*
$$p_1 \otimes p_2 \sqsubseteq q_1 \otimes p_2 \sqsubseteq q_1 \otimes q_2$$

*ii) immediately results from the idempondency of $\otimes$, by the substitution of $q_1$, respectively $q_2$ with q and applying the relation i):*
$$p_1 \sqsubseteq q \land p_2 \sqsubseteq q \Rightarrow p_1 \otimes p_2 \sqsubseteq q \otimes q = q \quad \blacksquare$$

*Proposition 1*, together with the transitivity of the refination and the commutativity of the product, enables the modular and hierarchical verification. The problem is to determine if $p \sqsubseteq q$, where $p$ is a specification and $q$ an implementation. The idea is to determine a chain of intermediate specifications $t_0, t_1,...,t_n$ so that $t_0=p$ şi $t_n=q$.

The intermediate specifications (including $p$ and $q)$ may be broken into components:

$$t_i = a_1 \otimes a_2 \otimes \ldots$$
$$t_{i+1} = (b_{11} \otimes b_{12} \otimes \ldots) \otimes (b_{21} \otimes b_{22} \otimes \ldots) \otimes \ldots$$

Then, we verify for each $j$, that:

$$a_j \sqsubseteq b_{j1} \otimes b_{j2} \otimes \ldots$$

From the monotony of the product $\otimes$ comparing with $\sqsubseteq$, it follows:

$$t_i \sqsubseteq t_{i+1}, \quad i \in \{0,1,...,n-1\}$$

and from the transitivity we establish that for $p \sqsubseteq q$.

If we also consider the property of idempotency the consecutive specifications can be partially covered: the refination between $p$ and q can be checked by breaking $p$ in more parts:

$$p = p_1 \otimes p_2 \otimes \ldots$$

and, then, by comparing each part with $q$. The parts of $p$ can be considered the properties that must be individually verified. If for each index $i$, $p_i \sqsubseteq q$, then $p \sqsubseteq q$.

It is obvious that the technique of modular and hierarchical verification, with a finite number of levels of specifications and with a finite number of components at each level is justified by *corollary 2*.

An alternative definition of the refination is to say that an "implementation" $q$ relatively is correct to a "specification" $p$, if $q$ operates properly in the environment of $p$. The question is whether this alternative definition is equivalent to the definition 8 (*definition 8)* of paper [LL09].

218

The following proposition answers positively to this question and therefore it connects the notions of absolute and relative correctness (see [LL09]).

**Theorem 1**. *Let us have two fuzzy processes p and q over the set of executions E,*

$$p \sqsubseteq q \iff \neg p \Leftrightarrow q \in R_E$$

**Demonstration.** *Let us have* $p = (\Delta^p_{X_p}, \Gamma^p_{Y_p})$ *and* $q = (\Delta^q_{X_q}, \Gamma^q_{Y_q})$ *(see figure1)*

$$\neg p \otimes q \in R_E \iff \Gamma^{p \otimes q} = 1_E \iff$$
$$(Y_{\neg p} \cap Y_q) \cup (\tilde{X}_{\neg p} \cap \tilde{Y}_q) \cup (\tilde{Y}_{\neg p} \cap \tilde{X}_q) = E$$
$$(X_p \cap Y_q) \cup (\tilde{Y}_p \cap \tilde{Y}_q) \cup (\tilde{X}_p \cap \tilde{X}_q) = E$$

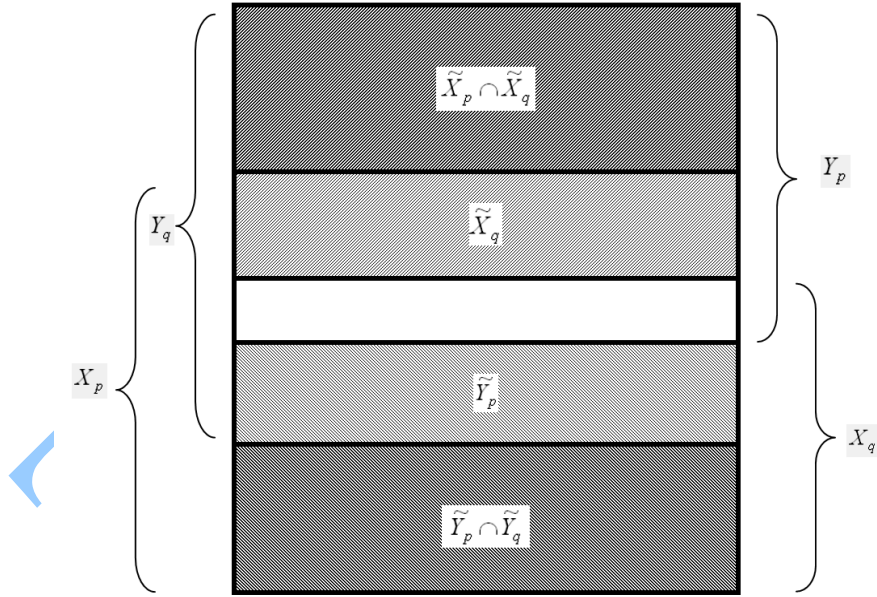

*Figure 1.* ∎

The above theorem allows us to verify whether an implementation satisfies the specification, by placing the implementation in the environment of the specification and then verifying the condition of the absolute correctness of their product. Our result is identical to that obtained in the classical approaches (i.e, [Ver94]).

219

We can give an alternative definition for refination in terms of testing: *q* is "better than or as good" as *p* if *q* passes all the tests that *p* can pass. Passing a test *r* can be seen as an absence of rejections when the device is connected to *r*.

The following theorem shows that this definition of refination is equivalent to the definition 8 from [LD01], and therefore it provides a new connection between the notions of absolute and relative correctness from the space of the fuzzy process.

**Theorem 2**: *Let us have two fuzzy processes p and q over the set of executions E,*

$$p \sqsubseteq q \iff \forall r : (r \otimes p \in R_E \Rightarrow r \otimes q \in R_E)$$

***Demonstration***. *From theorem 1 and proposition 1 we have that:*

$$r \otimes p \in R_E \Rightarrow -r \sqsubseteq p$$
$$r \otimes q \in R_E \Rightarrow -r \sqsubseteq q$$

*So, it is sufficient to show that:*

$$p \sqsubseteq q \iff \forall r : (-r \sqsubseteq p \Rightarrow -r \sqsubseteq q)$$

*The first implication follows from the transitivity of the refineries:*

$$p \sqsubseteq q \land -r \sqsubseteq p \Rightarrow -r \sqsubseteq q$$

*Reciprocally, let us have r = -p, then -r = p and from the reflexivity of the refination ⇒ -r ⊑ q. From the hypothesis -r ⊑ q and because -r = p, it follows that p ⊑ q* ∎

## 2 Robust processes

In concurrency theory it is often used "the testing paradigm", which we formulate in terms of fuzzy processes: being given a process *p*, which represents a known specification and a process *q*, which is part of a known implementation, and a process *r*, which represents the unknown part of the implementation, then

$$p \sqsubseteq q \otimes r$$

often called "the design inequality".

The following theorem solves this design inequality, characterizing its solutions as those fuzzy processes that refine a minimal solution.

**Theorem 3:** *Let us have the fuzzy processes p, q and r over the same set of executions E,*

$$p \sqsubseteq q \otimes r \Leftrightarrow p \oplus \text{-} q \sqsubseteq r$$

***Demonstration.*** $\quad p \sqsubseteq q \otimes r \Leftrightarrow \text{-} p \otimes (q \otimes r) \in R_E \Leftrightarrow$
$\text{-}p \otimes q \otimes r \in R_E \Leftrightarrow \text{-}p \otimes (\text{-} \text{-}q) \otimes r \in R_E \Leftrightarrow$
$\text{-}(p \oplus \text{-}q) \otimes r \in R_E \Leftrightarrow p \oplus \text{-}q \sqsubseteq r$ ∎

A classic design inequality is the software designing for embedded systems. In this sense, *p* is the specification known as embedded system, *q* is the known description of the underlying machine, and *r* is the unknown specification for software.

Often in the designing of systems it is expected that the subsystems (parts) are very easy to handle and have a defined behavior in any environment. Such features are modeled in the space of fuzzy processes using the properties of ***robustness***: the device specified by a fuzzy process accepts any execution, regardless the way in which the environment behaves towards the execution. Consequently, the environments (users) should be assumed to be completely unpredictable in the sense that they do not offer any guarantee in terms of avoided executions.

These observations support the following theorem, which shows how a fuzzy process can be "split" in two parts, a robust and a chaotic one.

**Theorem 4:** *Let it be a fuzzy process p over the set of executions E:*

  *i)* $p \sqcup \Omega \in R_E$
  *ii)* $p \sqcap \Omega \in H_E$
  *iii)* $p = (p \sqcup \Omega) \otimes (p \sqcap \Omega)$

***Demonstration.***

i) $\Delta^{P \sqcup \Omega}(x) = \min_{x \in X_p \cap E} \{\Delta^p_{X_p}(x), 1_E(x)\} = \Delta^p_{X_p}(x)$

$\Gamma^{P \sqcup \Omega}(x) = \max_{x \in Y_p \cap E} \{\Gamma^p_{Y_p}(x), 1_E(x)\} = 1_E$

then $p \sqcup \Omega \in R_E$

ii) $\Delta^{P \sqcap \Omega}(x) = \max_{x \in X_p \cap E} \{\Delta^p_{X_p}(x), 1_E(x)\} = 1_E(x)$

221

$$\Gamma^{p \sqcap \Omega}(x) = \min_{x \in Y_p \cap E} \{\Gamma_{Y_p}^p(x), 1_E(x)\} = \Gamma_{Y_p}^p$$

then $p \sqcap \Omega \in H_E$

iii) $q = (p \sqcup \Omega) \otimes (p \sqcap \Omega) = (\Delta_{X_p}^p, 1_E) \otimes (1_E, \Gamma_{Y_p}^p) = (\Delta_{X_p}^p, \Gamma_{Y_p}^p) = p$,

because $\tilde{X}_p \subseteq Y_p$ ∎

The robust fuzzy processes can be viewed as pure guarantees and the chaotic fuzzy processes can be seen as pure requirements. *Theorem 4* shows exactly the fact that any fuzzy process is the product resulted from a pure warranty and a pure requirement. Moreover, it gives a method to calculate the factors.

The next proposition shows that the product of two robust devices or of two robust environments is also robust, namely the fact that if all components of the system are robust, then the system is robust. Moreover, it indicates even other properties of closure for many fuzzy robust processes, in the finite case (comparing to the operations defined in [LL09]).

**Proposition 2**: *$R_E$ set is closed to $\otimes$, $\oplus$, $\sqcup$ and $\sqcap$.*

***Demonstration.*** *Let p and q be two fuzzy robust processes. Then the proposition is immediate if we calculate the sets of acceptable executions for*

$$p \otimes q, \; p \oplus q, \; p \sqcup q, \; p \sqcap q$$

*For example:*
$$Y_{p \otimes q} = (Y_p \cap Y_q) \cup (\tilde{X}_p \cap \tilde{Y}_q) \cup (\tilde{Y}_p \cap \tilde{X}_q) = E \quad \blacksquare$$

Using the properties of distributivity, commutativity, idempotency, etc. of the reunion and classic intersection, we obtain the following property:

**Proposition 3:** *Let us have three fuzzy processes p, q and r over the set of executions E,*

    i)     $p \sqsubseteq q \Leftrightarrow p \sqcup q = q \Leftrightarrow p \sqcap q = p$

    ii)    $p \sqcap (q \sqcup r) = (p \sqcap q) \sqcup (p \sqcap r)$

    iii)   $p \sqcup (q \sqcap r) = (p \sqcup q) \sqcap (p \sqcup r)$   ∎

Moreover, we notice that $\Omega \sqcup -\Omega = \Omega$.

**References**

[LD01]   L. Luca, I. Despi - *Toward a Definition of Fuzzy Processes*, Proceedings of the 5th International Symposium on Economics Informatics, Bucharest, pp. 855-859, May 2001.

[LL09]   L. Luca, L. L. Luca – *About Operations of Fuzzy Processes*, Proceedings of the 5th International Symposium, Timişoara, May 2009.

[Luc03]  Luca, L. – *Spaţii de procese fuzzy*, Editura Mirton, Timişoara, 2003

[Neg95]  R. Negulescu - *Process spaces*, Technical Report CS-95-48, Department of Computer Science, University of Waterloo, Ontario, Canada, December, 1995.

[Neg98]  R. Negulescu - *Process Spaces and Formal Verification of Asynchronous Circuits*, PhD thesis, Department of Computer Science, University of Waterloo, Ontario, Canada, August, 1998.

[Ver94]  T. Verhoeff - *The testing paradigm applied to network structure*, Computing Science Notes 94/10, Department of Mathematics and Computer Science, Eindhoven University of Technology, The Nederlands, 1994.